\documentclass[doublespacing]{elsart}
\usepackage{icarus}
\usepackage{amssymb}
\usepackage{ulem}
\usepackage{natbib}

\bibpunct{(}{)}{;}{a}{,}{,}

\usepackage{graphicx}
\newcommand{\BibTeX}{ \textrm{B\kern-.05em\textsc{i\kern-.025em b}\kern-.08em
    T\kern-.1667em\lower.7ex\hbox{E}\kern-.125emX} }
    
\newcommand{\cACB}     [1]{{\bf #1}}

\begin{document}

\begin{frontmatter}



\title{Clumps in the Outer Disk by Disk Instability: Why They are Initially Gas Giants and the Legacy of Disruption}


\author[acb1,acb2]{Aaron C.~Boley}, 
\author[th]{Tristen Hayfield},
\author[lm]{Lucio Mayer}, and
\author[rhd]{Richard H.~Durisen}

\address[acb1]{Institute for Theoretical Physics, University of Zurich, Winterthurerstrasse 190,
Zurich, CH-8057, Switzerland}
\address[acb2]{Astronomy Department, University of Florida,
211 Bryant Space Science Center,
PO Box 112055, USA}
\address[th]{Physics Department, ETH Z\"urich, Z\"urich,
CH--8093 Z\"urich, Switzerland}
\address[lm]{Institute for Theoretical Physics, University of Zurich, Winterthurerstrasse 190,
Zurich, CH-8057, Switzerland}
\address[rhd]{Department of Astronomy, Indiana University,
727 East 3rd Street, Bloomington, IN 47405, USA}

%% This copyright statement isn't required at any stage by the Icarus
%% Editorial Office or Elsevier.  However, until you sign over the
%% copyright to Elsevier prior to publication (or negotiate with them
%% about copyright), you own the copyright to anything you create.
%% Just to keep things unambiguous, always include a copyright statement
%% or explicitly dedicate your work to the public domain.

%% ----- ELSEVIER STUFF -----
%% The commands below up to the \end{frontmatter} are commented out
%% so that we can do some Icarus-required formatting on the second and
%% third pages that is not required later on by Elsevier.  So when
%% your paper gets accepted, and you are ready to start dealing with
%% Elsevier, copy your abstract and keywords up here, uncomment these
%% lines, and comment out the ICARUS STUFF below.
%% 
%% Alternately, you might just want to move these abstract, keyword,
%% and end frontmatter commands down, and comment out the ICARUS STUFF
%% commands.  It doesn't matter.

% \begin{abstract}
% % Text of abstract
% 
% \end{abstract}
% 
% \begin{keyword}
% % keywords here, in the form: keyword \sep keyword
% 
% 
% % PACS codes here, in the form: \PACS code \sep code
% 
% \end{keyword}

%% ----- END ELSEVIER STUFF -----

\end{frontmatter}

%% ----- ICARUS STUFF -----
%% Some formatting on the first, second, and third pages are required
%% by the Icarus Editorial Office that are not required by Elsevier.
%% This section contains those things.  When you are ready to transition
%% to ``Elsevier'' mode, copy your abstract and keywords up into
%% the ELSEVIER STUFF section, and then you can just delete everything
%% in this section.

%% We need to list the number of manuscript pages, figures, and tables. 
%%
%% Rather than manually count these things out, we'll use a little
%% trick here from Paul.  All you have to do is place three \label{}
%% tags on the last page, the last table, and the last figure, that
%% way these values are automatically updated (as long as you remember
%% to move the lasttable and lastfig labels when you add or remove
%% tables and figures).

\begin{flushleft}
\vspace{1cm}
Number of pages: 38 \\
Number of tables: 1\\
Number of figures: 8\\
\end{flushleft}

%% Don't worry about finding the various last* tags and deleting them
%% when you go to ``Elsevier'' mode if you don't want to, they should be
%% silently ignored.

%% The second page should indicate a proposed running head of not more 
%% than 55 characters, and the name and address to which editorial 
%% correspondence and proofs should be directed.  The pagetwo 
%% environment that icarus.sty provides will make page two for you,
%% just give the running head as an argument to the environment, and
%% then your correspondence address inside.
\begin{pagetwo}{Legacy of Clump Disruption     }
%                        1         2         3         4         5
%               1234567890123456789012345678901234567890123456789012345

Aaron C.~Boley\\
Department of Astronomy\\
University of Florida\\
211 Bryant Space Science Center\\
PO Box 112055\\
Gainesville, FL, 32611-2055, USA\\
Phone: (352) 392-2052\\
Email: aaron.boley@gmail.com

\end{pagetwo}

\begin{abstract}

We explore the initial conditions for fragments in the extended regions ($r\gtrsim 50$ AU) of gravitationally unstable disks.  We combine analytic estimates for the fragmentation of spiral arms with 3D SPH simulations to show that initial fragment masses are in the gas giant regime.  These initial fragments will have substantial angular momentum, and should form disks with radii of a few AU.  We show that clumps will survive for multiple orbits before they undergo a second, rapid collapse due to H$_2$ dissociation and that it is possible to destroy bound clumps by transporting them into the inner disk.  The consequences of disrupted clumps for planet formation, dust processing, and disk evolution are discussed.  We argue that it is possible to produce Earth-mass cores in the outer disk during the earliest phases of disk evolution.
\end{abstract}

% %% Keywords should appear after the abstract. 

%% ----- END ICARUS STUFF -----

\section{Introduction}

Protoplanetary disks are likely to be massive during their initial phases of evolution. Collapse calculations demonstrate that gravitationally unstable disks do form \citep[e.g.,][]{vorobyov_basu2009} and that their evolution includes phases of strong gravitational instability \citep[e.g.,][]{vorobyov_basu2006}.  In the inner disk ($r\lesssim 50$ AU), cooling times relative to local dynamical times are too long for the instability to result in fragmentation, and the disk reaches a self-regulating state \citep[e.g.,][]{boley_etal_2006,durisen_etal_2007}.   There may be some exceptions due to changes in disk chemistry \citep{mayer_etal_2007}, but these situations require further study.  In contrast, fragmentation in the outer, extended disk ($r \gtrsim 50$ AU) becomes quite possible if the \cite{toomre1964} $Q$ can be driven toward unity by, e.g., mass loading \citep{boley2009}.   Whether these fragments typically produce brown dwarfs \citep{stamatellos_etal_2007,stamatellos_whitworth2009} or gas giant planets is a topic of debate.  However, as we argue here, clumps that become destroyed can be just as important to planet formation and disk evolution as clumps that remain bound.   

In this paper, we study disk fragmentation conditions and make estimates for initial fragment masses; we describe the initial angular momenta of fragments; and we discuss the consequences of clump disruption for planet formation. We present simulation data and describe toy models in section 2.  In section 3, we derive proper estimates for fragment masses, and find consistency between our estimates and simulation data.  We estimate the expected initial angular momenta of clumps in section 4, which are also in rough agreement with simulation data. We use a polytropic model in section 5 to estimate the time during which  a clump could be tidally disrupted, including accretion effects.  These sections culminate to show that clump disruption is a real possibility during disk evolution.  For section 6, we speculate on the consequences of disrupted clumps for dust processing, core formation, and the FU Orionis phenomenon.  Our conclusions are given in section 7.

\section{Models}

We present two parameterized models.  Model A  is used to show what we expect for fragmentation around A stars.  We assume a  temperature profile $T=350 (1 AU/r)^{1/2} + 10 $ K, a mean molecular weight $\mu=2.3$, and the epicyclic frequency $\kappa\approx\Omega\approx \Omega_{\rm Keplerian}$ for central star $M=1.5 M_{\odot}$.   The surface density $\Sigma$ is given by the \cite{toomre1964} parameter $Q= c_s\kappa/(\pi G\Sigma)$, where $c_s$ is the local sound speed.   The entire disk does not need to be gravitationally unstable \citep[$Q\lesssim 1.7$,][]{durisen_etal_2007}, and the total disk mass for the model could vary considerably depending on the size of the low-$Q$ region.  If  Model A's disk has a $Q=1.5$ between $r\sim 100$ and 150 AU, the mass in this region would be about 0.15 $M_{\odot}$.   Our second model, Model M, represents a disk orbiting an M star, embedded in an envelope with a 30 K irradiation temperature.  For $Q=1.5$, the mass contained between $r\sim 100$ and 150 AU is 0.04 M${\odot}$.  In addition to these parameterized models, we present a simulation with initial conditions (ICs) based on the simualtion ``SIMA'' from \cite{boley2009} just before the disk fragments.  We refer to this model for the rest of the paper as SPHSIM to avoid confusion with SIMA and the analytic models described above.  In SPHSIM, a $0.3\rm M_{\odot}$ star is surrounded by an $ r\sim 400$ AU disk that is accreting mass from an envisaged envelope at $\sim 10^{-5}\rm M_{\odot}~yr^{-1}$.  This accretion rate is consistent with what one expects in a protostellar collapse with a background temperature of 30 K.   At the time our simulation begins, the disk has a mass of $\sim 0.19\rm M_{\odot}$. The inner radius is  at $r\sim 18$ AU and an outer radius of $r\sim 510$ AU.   We do not include additional mass loading,  so SPHSIM represents a last-burst scenario, i.e., the final phase of fragmentation that this disk is likely to experience.  The data were interpolated from the CHYMERA \citep{boley2007} cylindrical grid to an SPH realization via a density-weighted Monte Carlo sampling.  

SPHSIM is run using GASOLINE \citep{gasoline}, which is a multipurpose code designed to model structure at various scales, e.g., planetary disks as well as cosmological structure formation.
 One million particles are followed,  with a particle mass $\sim 2\times 10^{-4} M_J$.  The spline force softening is set to 2 AU for the star and 0.5 AU for all other particles.   The radiative cooling algorithm is the same as that described in \cite{boley2009}, but adapted to SPH.  The cooling is calculated from $\nabla\cdot F = -(36\pi)^{1/3} s^{-1}\sigma (T^4-T_{\rm irr}^4) (\Delta \tau + 1/\Delta \tau)^{-1}$, where $s=(m/\rho)^{1/3}$ and $\Delta \tau=s\kappa\rho$ for the local opacity $\kappa$, particle mass $m$, and density $\rho$. The factor $36 \pi$ comes from defining the radius of a resolution element as $r=(3 m /(4\pi \rho))^{1/3}$ and setting the ratio between the radiative flux and the divergence of the fux to be the area over the volume of the resolution element. \cite{dalessio_etal_2001} opacities are used, with a $1~\mu\rm m$ maximum grain size.  The irradiation temperature $T_{\rm irr}=30$ K everywhere.  This cooling approximation is good for the outer disk regions, where midplane optical depths are $\lesssim 1$.   The disadvantage to this approximation is that it neglects the effects of radiation from a collapsing clump on its surrounding medium.  

 In addition to the radiative cooling algorithm described above, GASOLINE has been augmented with the {\citet{read_etal_2009}} OSPH modifications in order to address the SPH limitations outlined, most recently, by {\citet{agertz_etal_2007}}. We find that OSPH is superior to other suggested modifications {\citep[e.g.,][]{price_2008}} because (1) OSPH works for a self-gravitating fluid, (2) recovers the correct timescale for the growth of the Kelvin-Helmholtz instability without introducing new free parameters, and (3) allows for convergence testing of the hydrodynamics by increasing the number of neighbors without being compromised by the tensile instability. We also note that GASOLINE uses a fixed gravitational softening, while the SPH smoothing length $h$ is variable, defined by 32 neighbors.  Using a variable softening length can induce fluctuations in the
potential energy of particles, which inevitably leads to errors in energy conservation. On the other hand, Nelson (2006) showed that fixed softening could lead either to either enhanced or suppressed clumping.   Although there are ways of improving
energy conservation that would permit the use of variable softening lengths {\citep{price_monaghan_2007}},
this has yet to be implemented in GASOLINE and represents a future code development project.  Instead, we have chosen to ensure that energy is conserved and set the mass resolution such that the gravitational softening length is larger than the typical SPH smoothing length in dense structure, e.g., spiral arms {\citep[see discussion in][]{mayer_etal_2004}}.  The fragments that form in SPHSIM (below) have a median softening to smoothing length ratio of about 4, so SPHSIM should err on the side of slowing the collapse of fragments.  Moreover, GASOLINE showed highly satisfactory results in the Wengen Test 4 comparison project, where a highly unstable disk was followed in detail by several codes.  In particular, Gawryszczak \& Mayer (2008) reported strikingly good agreement between GASOLINE and the FLASH AMR code when following fragmentation in a self-gravitating disk.

Three clumps form in SPHSIM (C1, C2, and C3), and their initial properties are listed in Table 1. We computed the number of particles per Jeans mass according to $M_{\rm Jeans}/M_{\rm particle}=2.92 c_s^3/(M_{\rm particle} G^{3/2}\rho_{\rm peak}^{1/2})$, for sound speed $c_s$ and peak density $\rho_{\rm peak}$ in the fragmenting spiral arm.  Fragmentation is numerically well-resolved, with $M_{\rm Jeans}/M_{\rm particle}\sim 2000$, satisfying the \citet{bate_burkert_1997} criterion \citep[see also][]{nelson2006}.   Particles are flagged as fragment members using SKID \citep{stadel2001}, which groups particles according to density gradients and then completes an iterative unbind for each particle in the group.     When comparing the results to SIMA, it should be noted that Boley (2009) reported a clump mass of 20 M$_J$ for the end of the simulation. This estimate included mass growth for about 1.75 orbits, so it  does not represent the clump's initial mass, which is between 4 and 5 M$_J$.  Boley expressed reservation in the Letter about accepting the final masses from his simulations because, e.g., the radiative effects of the clump on its surroundings were not modeled, and the resolution was too low to follow the evolution of the clump itself.  In addition, SIMA only forms one clump, while SPHSIM forms three.  Although integrated or time-averaged quantities between two realizations of a simulation should give comparable answers, detailed structure, especially clumps, are extremely sensitive to initial conditions (see Wengen 4 Comparison Project\footnote{www.astrosim.net}).  The difference in the number of fragments between SPHSIM and SIMA are not considered by us to be failures of either model.

\section{Clump Mass}

In this section, we calculate initial clump masses for unstable disks.  {\it Why do we care about initial clump masses?} Although clumps are likely to accrete, the evolution of the system will depend on its initial state.   For example, a non-rotating, one Jupiter-mass clump will contract for about a few$\times10^5$ yr, until molecular hydrogen dissociates, causing rapid collapse \citep[e.g.,][]{tohline2002}. We refer to this contraction timescale as the clump's primary contraction time ($\tau_1$).   For five Jupiter masses, $\tau_1$ is reduced to a few $\times10^4$ yr, and for an initial mass of ten Jupiter masses, the clump will collapse in $<10^4$ yr \citep{helled_bodenheimer_2009}.  A factor of ten in the initial mass can change the clump's initial evolution timescale by a factor of  about 100.  The spatial scale of fragmentation, which is related to the initial mass, also determines the rotational angular momentum (next section).  Once the core of the clump collapses, this angular momentum barrier should lead to the formation of a circumplanetary/brown dwarf disk.  This disk is expected to control the long-term accretion history of the clump, as material entering the Hill sphere of the collapsed planet will have angular momentum from disk shear.  In order to avoid confusion between fragmentation, i.e., the formation of the clump, and the rapid collapse that follows dissociation of H$_2$, we refer to the latter as dissociative collapse.

From a practical standpoint, initial masses can be well-constrained using global simulations, while the subsequent evolution can only be modeled poorly at this time.  Simulating clump growth requires resolving convection, the photosphere,  chemical changes, the clump's effect on the surrounding medium, the core/disk transition region, and the gas flow into the Hill sphere.  Even at our resolution of about 5000 particles per Jupiter mass, this is a daunting task, and best addressed by high-resolution simulations of individual clumps.  For all of these reasons, we argue that constraining initial clump masses is fundamental to understanding the fragmentation process in disks and gas giant planet formation.

{\it What do we expect for initial clump masses?}  In order to calculate the fragment mass, we need to know the local surface density and size scale for fragmentation. A back-of-the-envelope estimate is the Toomre mass 
\begin{equation}
M_T=\pi (\lambda_T/2)^2\Sigma
\end{equation}
 \citep[e.g.,][]{nelson2006}, where the Toomre wavelength, 
 \begin{equation}
 \lambda_T=2 c_s^2/(G\Sigma), 
 \end{equation}
is roughly the most unstable {\it radial} wavelength for local sound speed $c_s$ and smooth surface density $\Sigma$.  The surface density can be calculated  from the Toomre parameter using $\Sigma=c_s\Omega/(\pi G \left<Q\right>)$, and this value can be used to find the Toomre mass for a given $c_s$ and $\left<Q\right>$, where we have taken $\kappa\approx\Omega$.  The brackets are used to denote that $\left<Q\right>$ is a smooth, axisymmetric quantity.  We can rewrite the Toomre wavelength 
\begin{equation}
\lambda_T= 2\pi \left<Q\right> f_{\rm e} f_{\rm g} H,
\end{equation}
 where the local scale height 
 \begin{equation}
 H=c_s/(f_e f_g \Omega),
 \end{equation}
and $f_e$ and $f_g$ are shape 
factors of order unity that depend on the equation of state and on self-gravity effects, respectively.  Setting $\left<Q\right>=1$ and calculating a Toomre mass based on the most unstable radial wavelength for the unperturbed axisymmetric disk ($\lambda_T$) includes material over a radial extent that is $2\pi H$.    Gravitational instabilities can produce strong spiral waves and local density perturbations, making such use of the axisymmetric measure $\lambda_T$ over large \cACB{radial} scales in the non-linear regime, where fragmentation occurs, extremely dubious.  The Toomre mass is strictly an estimate for the disk mass that becomes incorporated into one wavelength of the resulting spiral waves.  Fragmentation is best described in the context of spiral arms.

Instead of using measurements that correspond to an axisymmetric disk, we use length scales and  surface density perturbations that are appropriate for spiral arms. The radial extent of fragmentation can be estimated using the results of \citet[][hereafter DHP2008]{durisen_etal_2008}, who used the virial theorem to show that a disk, under isothermal conditions, is most susceptible to fragmentation within a region $\delta r$ from the corotation of a spiral wave.  They demonstrated this behavior using isothermal hydrodynamics simulations.  Other studies have confirmed that fragments tend to form near the corotation of spiral waves, even when radiative physics is included \citep{boley_durisen2008,boley2009}.  DHP2008 found that an isothermal spiral shock, with a corotation at $r$, is stable against fragmentation for
\begin{equation}
\left(\frac{\delta r}{r}\right)^2>\frac{4 \pi^2  f_{\rm DHP}^2 f_g}{81(\sin i)^2\left<Q\right>m^2},
\end{equation}
where $m$ is the number of spiral arms and $i$ is the pitch angle of the spiral, which is typically $i\approx 10^{\circ}$ in gravitationally unstable disks \citep[see][]{boley_durisen2008,cossins_etal_2009}. Using the DHP2008 definition for $f_{\rm DHP}$, we find that $f_{\rm DHP}=H\mathcal{M}^2m/(\pi r)$, where $\mathcal{M}$ is the Mach number for the isothermal shock.  Strictly, we are using $\mathcal{M}$ to indicate the density enhancement in the spiral arm over the smooth distribution, and this should be kept in mind when we quantify our results.  If the shock truly is isothermal, then the Toomre wavelength within the overdensity  will be 
\begin{equation}
\lambda_T'=2c_s^2/(G\Sigma\mathcal{M}^2)=\lambda_T/ \mathcal{M}^{2}.
\end{equation}
 We expect fragmentation to occur only when $\lambda_T'$ becomes equal to $2\delta r$.  The ratio of these quantities is
\begin{equation}
\left(2\frac{\delta r\mathcal{M}^2}{\lambda_T}\right)^2=\frac{132 f_g^3}{81\pi^2 \left<Q\right>^3}.
\end{equation}
Fragmentation should occur at corotation when 
\begin{equation}
f_g\approx 1.8\left<Q\right>.
\end{equation}

Now that we have an estimate for the location and radial extent of fragmentation, we need an expression for the surface density perturbation, relative to the axisymmetric density, that will permit fragmentation. We also need to know the azimuthal width of the shock, which can be estimated by assuming it is similar to the full height ($2H$) of the disk. For estimating  $H$ within a spiral arm, we refer to {\citet{boley_durisen2006}}, who showed that the scale height in the post-shock region of a self-gravitating, isothermal spiral shock should be reduced by the factor 
\begin{equation}
F({\mathcal{M}})=\left(\frac{q+\mathcal{M}^2}{q+1}\right)^{1/2}
\end{equation}
 due to the increased gas density, where $q\equiv$ external gravity/self-gravity for the axisymmetric disk.  The self-gravity of a disk at its scale height is well approximated by $2\pi G \Sigma$, while the star's gravity by $\Omega c_s$. Combining these terms yields  
 \begin{equation}
 q\approx \left<Q\right>/2.
 \end{equation}
  However, the self-gravity of the unperturbed disk should be included in our definition of $f_g$ as well.  We approximate this effect by setting 
 \begin{equation}
  f_g\approx F(\mathcal{M}/q^{1/2})=\left(\frac{\left<Q\right>+4\mathcal{M}^2/\left<Q\right>}{\left<Q\right>+2}\right)^{1/2}
  \end{equation}
 for $\left<Q\right><2$ and $f_g=F(\mathcal{M})$ otherwise.   Multiple simulations have shown that an initial $\left<Q\right>\lesssim 1.4$ is required for an isothermal disk to fragment \citep[e.g.,][]{tomley_etal_1994, nelson_etal_1998,johnson_gammie2003,mayer_etal_2004, durisen_etal_2007}. Using $\left<Q\right>=1.4$ in equation (8) gives $f_g \approx 2.5$, which corresponds to $\mathcal{M}\approx2.7$ in equation (11).  As $\left<Q\right>$ is lowered, the density perturbation that is required to induce fragmentation is also lowered.

%Strictly, $f_g$ should also take into account the self-gravity of the unperturbed disk.  However, in the limit that self-gravity is completely dominant, the hydrostatic density structure of a disk has a surface at $\zeta\approx\left<Q\right> c_s/Omega$. In the non-self-gravitiating case, about 70\% of the mass is contained within the disk's scale height. Using this mass criterion to compare $h$ with $\zeta$, we find $h=0.7 \left<Q\right> c_s/\Omega$.

Now that we have estimates for the radial extent of the fragmenting region ($\lambda_T'$), for the width of the spiral ($2H$, using $f_g$ and equation 4), and for the surface density perturbation over the axisymmetric disk ($\Sigma \mathcal{M}^2$), we can find the initial clump mass within the context of a fragmenting spiral arm:
\begin{equation}
 M_f=2 \lambda_T \frac{\Sigma c_s}{\Omega f_g}.
 \end{equation}
Figure 1 shows the initial mass if it were calculated assuming $M_T=\pi \Sigma (\lambda_T/2)^2$, 
assuming the Model A temperature profile, and by using our estimate for $M_f$; we assumed $\left<Q\right>=1.4$.  The curve for Model M shows what we expect from the model parameters, and the actual clump masses are shown by symbols.  C1 and C2 match our $M_f$ estimate well, and the initial masses are in the gas giant regime.  C3 also has a mass in the gas giant range, but our $M_f$ calculation is an overestimate.  This may be due to differences between prompt and delayed fragmentation (see DHP2008 for a detailed discussion).  C1 and C2 form near corotation (prompt fragmentation), while C3 appears to form during the collision between the wake of C1 and an outer arm (delayed fragmentation).    In contrast to our $M_f$ estimates, $M_T$ indicates that {\it initial} masses should be in the range of brown dwarfs for $r\gtrsim 80$ AU, even for $\left<Q\right>=1$.  This has led to recent claims \citep{kratter_etal_2009} that,  even initially,  gas giant-mass clumps should be atypical, which is inconsistent with our analysis and our simulation data.  Figure 2 shows a close-up snapshot of C1  before and just after fragmentation.  Fragmentation is clearly confined to the spiral arm.  The pre-fragment material has an $H\approx 1.2$ AU and a $\delta r\approx 9$ AU.  For comparison, we estimate that $H$ should be $\approx2.2$ AU and $\delta r\approx8.5$ AU using our analysis above.

\section{Initial Angular Momentum}

The radial extent of fragmentation suggests that clumps should have substantial angular momentum from the shear in the disk.  The specific angular momentum of a newly-formed clump can be approximated by 
\begin{equation}
J_{\rm init}\approx 1/3(\Omega(r+\delta r)(r+\delta r)^2-\Omega(r)r^2),
\end{equation}
 which for $\delta r\ll r$, $J_{\rm init}
\approx 1/6 (G M_{\rm star}/r)^{1/2}\delta r$, where $\delta r = \lambda_T'/2$.  This estimate assumes that the difference between the orbital angular momentum of a clump's outermost material and of the centroid of fragmentation ($r$) goes entirely into rotation.    The factor of 1/3 comes from assuming we have a rigid rod of length $2\delta r$ for the moment of inertia, based on the shape of the collapsing region in the left panel of Figure 2.  Figure 3 shows the initial angular momenta for hypothetical fragments that we expect would form in Models A and M, as a function of radius, according to equation (13).   Clumps that form at $r\sim100$ AU should have  $J\sim\rm few\times10^{18} cm^2s^{-1}$, which is roughly consistent with the simulation data.   This also strongly suggests that  a circumplanetary disk can form with a radius of approximately an AU (Fig.~4) after subsequent clump contraction.  The expected angular momentum radius, i.e., where we expect rotation to limit further contraction, can be estimated by
\begin{equation}
r_J\approx \frac{\delta r^2}{36r}\frac{M}{M_f}.
\end{equation}
  Using our estimate for  the fragmentation mass $M_f$ from section 3, this becomes 
  \begin{equation}
  r_J\approx  \frac{\pi^2 \left<Q\right>^2 v_K f_g r}{144 c_s \mathcal{M}^4},
  \end{equation}
  where $v_K$ is the Keplerian orbital speed.  The sizes of the clumps in the SPHSIM (Table 1) are consistent with this estimate (Fig.~4), even though we overestimate $J$ by a factor of about two.  This suggests that thermal pressure is still an important component in a clump's initial size.  Indeed, the initial $T/|W|$ for the clumps is roughly 0.2, where $T$ is the total rotational energy and $W$ is its potential energy.  These clumps are initially more like rapidly rotating spheroids than true central object+disk systems.  During the contraction phase, the clump may become susceptible to dynamical instabilities, for example, the bar instability \citep[e.g.,][]{durisen_etal_1986}.  Such a dynamic event would rapidly rearrange the angular momentum distribution of the object, leading to rapid outward transfer of angular momentum and overall expansion of the clump.  Convection could also play a role in redistributing angular momentum, but its overall effect is uncertain.    Unless some mechanism can transfer angular momentum inward from the outer mass shells to the central regions, our estimate should be valid, and shows that the high-$J$ material in the clump will be unable to collapse to the size scales of Jupiter.  Once dissociative collapse is reached, a disk should form.  To illustrate this point further, we compare Jupter's rotational angular momentum to a wide-orbit clump's initial $J$.  It is unclear whether Jupiter's angular momentum is consistent with the $J$s for planets on wide orbits, but it provides a reference value.  For this estimate, we assume that Jupiter is a rigidly rotating sphere with a radius $\sim 7\times10^9$ cm and a rotational period of 10 hr.  This gives $J_{\rm Jupiter}\sim 3\times 10^{15}$ cm$^2$ s$^{-1}$, which is about three orders of magnitude smaller than a clump's initial $J$. 

We have ignored the possibility that a large fraction of the angular momentum given to the clump from shear goes into altering the orbital angular momentum of the  clump.  We expect for this to affect our result by a factor of order unity.  Likewise, the orientation of the arm just before fragmentation and the difference between the assumed moment of inertia (the rigid rod) and the actual one will also cause errors of order unity. Nevertheless, the SPHSIM data show that our estimate for $J$ is valid to an order of magnitude and provides a reasonable upper limit.

\section{Contraction Timescale}

So far, we have defined the mass scales and initial $J$ that we expect for fragmentation.  Whether these clumps can become bound objects depends, in part, on their  primary contraction timescale.  Once central temperatures reach $T_c\sim2000$ K, $H_2$ begins to dissociate, and a rapid collapse to sizes of a few Jupiter radii will follow because energy goes into dissociation instead of thermal support.  After the H$_2$ collapse, the clump will be stable against tidal disruption owing to its small size.   In contrast, a clump that is transported into the inner disk before it undergoes dissociative collapse could become tidally destroyed.  This is illustrated by Figure 5, which shows a snapshot of a clump being tidally disrupted in SPHSIM.    Two clumps interact, and C2 from Table 1 is put on an eccentric orbit.  As it approaches periastron, the clump's volume becomes much greater than its Hill volume, and it is destroyed.  The disruption of C2 takes place about 1500 yr after its formation.  Is this disruption physically motivated? Should the clump have already undergone dissociative collapse?  In this section, we determine whether the primary contraction timescale ($\tau_1$) is long enough to make clump disruption a real possibility in protoplanetary disks.

Helled (2009, private communication) has graciously shared her data with us \citep[from][]{helled_bodenheimer_2009}, which show that clumps with masses of a few $M_J$ roughly follow an $n=2.3$ polytrope. In order to calculate $\tau_1$ for a variety of conditions, we have developed a poor-man's gas giant evolution code using polytropes.  By assuming initial radii and masses as calculated above, an initial polytrope solution can be determined.  Using the polytrope profile and \cite{dalessio_etal_2001} Rosseland mean gray opacities, a photosphere can be calculated, which gives us the luminosity $L=4\pi R_{\rm eff}^2 \sigma T_{\rm eff}^4$. The time step between iterations is $\Delta t = 0.01 U/L$, where $U$ is the internal energy of the polytrope. For a given step, the total energy of the polytrope is updated according to $E= U+W -\Delta t L$.  The potential energy of the system is then calculated as $W=3E\frac{\gamma-1}{3\gamma-4}- \frac {3 G M\Delta M}{R(5-n)}$, where $\gamma=7/5$ is the thermodynamic adiabatic index, which is not necessarily the same as the structural adiabatic index $\gamma_P=1+1/n$.   The term with $\Delta M$ accounts for mass that is accreted over $\Delta t$.  Once the new potential energy is determined, a new radius can be calculated using $R=-3GM^2/((5-n)W)$, and the new internal energy can be calculated from the virial, i.e., $U=-W/(3(\gamma-1))$.  With the new mass and radius, the density and temperature profiles can be updated.  This continues until $T_c>2000$ K.  For all calculations here, $n=2.3$ and the maximum grain size assumed in the opacities is $1\mu m$.   Unfortunately, we ignore clump rotation for these calculations, even though, as discussed above, we expect it to influence subsequent clump evolution. To reiterate, this omission is expected to have two principal effects: (1) Rotation is extra support, and the contraction should be altered, especially when dynamical instabilities set in.  (2) A nonspherical contraction will likely alter the amount of material in the clump that is exposed to high temperatures, but our calculations present an order of magnitude estimate.

We now consider the structure and contraction of hypothetical clumps using the method described above.   Polytrope Clump 1 (PC1) is followed without accretion, and its mass is set to $3 M_J$ with an initial  polytrope radius $\sim 5$ AU.  The photosphere is initially at a radius $\sim 3$ AU, with a temperature of 22 K.  The background irradiation could affect the contraction timescale, but we do not address that detail here.    The central temperature and density evolution for PC1 is shown in the $\rho-T$ plane in Figure 6.  Symbols indicate $10^4$  yr intervals.   The contraction time for PC1 is $8\times10^4$ yr, which is very similar to the results of Helled \& Bodenheimer.  This timescale is about $100$ orbits at 100 AU for a 1.5 $M_{\odot}$ star, giving a large window of opportunity for, e.g., clump-wave and clump-clump interactions to transport clumps inward or outward.  However, we also need to address whether accretion will decrease the contraction timescale such that disruption becomes unlikely.

We assume that the clump is accreting at its maximum rate, where $\dot{M}$ onto the clump is limited solely by the rate that mass can be delivered to the planet's Hill sphere.  For simplicity, consider a cylinder centered on the fragment, with Hill radius $R_H$.  The flow of material into into this cylinder should be dominated by the shear flow around the fragment, which yields
\begin{equation}
\dot{M}=2 \int_0^{\pi/2}\frac{\Omega\left(r_0\right)}{2}\Sigma\left(r_0\right) R_H^2\cos i\sin i d i,
\end{equation}
where $r_0$ is the orbital radius and $i$ is the angle from the perpendicular to the radial direction. We have assumed that $\Sigma$, the surface density at $r_0$, is constant over perturbations of $R_H$ from $r_0$.  Evaluating the integral gives 
\begin{equation}
\dot{M}\approx 2\times 10^{-7}\ \rm M_{\odot}\ yr^{-1} \left(\frac{r_0}{100\ AU}\right)^{1/2}\left(\frac{M}{M_{J}}\right)^{2/3}\left(\frac{M_{star}}{M_{\odot}}\right)^{-1/6} \frac{\Sigma}{10\ g\ cm^{-2}},\end{equation}
for clump mass $M$. We have chosen to normalize the function to $\Sigma=10\rm~g~cm^{-2}$ because this surface density corresponds to an unstable disk with $M_{\rm star}= M_{\odot}$ and $T=10$ K at $r=100$ AU.  As the clump accretes, the size of its Hill sphere grows, allowing it to capture more mass.  As it becomes more massive, the evolution time toward rapid dissociative collapse decreases.  

Figure 7 shows the ratio of the clump mass just before $T_c=2000$ K to the initial mass, as a function of the initial fragment mass.  Although we expect initial clump masses to typically be a few M$_J$, we show higher masses for completeness.   Each fragment is assumed to grow at the rate given by equation (17), and has an initial polytrope radius of 5 AU.  For calculating $\dot{M}$, we assume a primary mass of 1 M$_{\odot}$, $\Sigma=10$ g cm$^{-2}$, and an $r_0=100$ AU. The accretion rate is updated every time step, allowing $\dot{M}$ to grow as the clump grows.  For comparison with PC1, a maximally accreting 1 M$_J$ clump (PC2) is shown on the $\rho$-$T$ plane along with PC1 (Fig.~6).  This clump grows to about  14 M$_J$ before it reaches dissociative collapse.  Recall that this mass actually represents a rapidly rotating clump, which should form a proto-gas giant/brown dwarf + disk system after dissociative collapse. The timescale to reach collapse is also shown in Figure 7, which indicates that fragments with initial masses less than about 6 M$_J$ remain susceptible to tidal disruption for $\sim10$ orbits at 100 AU around a 1.5 M$_{\odot}$ star.  As the initial fragment mass increases, accretion becomes less important to the precollapse evolution, but the time available for disruption continues to decrease.   

%Our simulation and calculation for $J_{\rm init}$ show that disks are expected to form around contracting clumps.  If we assume that mass accretion is controlled by the circumplanetary disk, we can estimate an average mass accretion rate for  a $1M_J$ planet based on an $\alpha$-model, where $\dot{M}\approx 3\alpha c_s^3/(GQ_{\rm disk})$.  Setting $\alpha=0.01$, $Q_{\rm disk}=1.4$, and $T=30K$ for $c_s$, we find that $\dot{M}\sim10^{-7} M_{\odot} \rm yr^{-1}$.

We do not claim that equation (17) represents the rate that a clump must accrete material; instead, we use it only as a rough upper limit for the accretion rate.  As discussed above, rotation could affect the contraction timescale and lead to dynamical instabilities in the rapidly rotating clump.  Additional caveats include the following:  (1) The material entering the clump's Hill sphere may have non-negligible thermal pressure. (2) The energy gained due to the accretion of gas may not be efficiently radiated away. (3) Some material entering the Hill sphere may be unable to shed enough angular momentum to become part of the clump. (4) Convection could rearrange the mass and angular momentum distribution of the clump, leading to additional instabilities.  It is unclear how these assumptions will affect out accretion estimate in detail, but we expect that these processes will decrease the average mass accretion rate of the clump, which will increase the time a clump remains susceptible to disruption. 
We conclude that the disruption of clumps during the earliest phases of disk evolution is quite possible.

\section{Consequences of Clump Disruption for Planet Formation}

The work in the previous sections demonstrates that clumps can remain in their primary contraction phase long enough after their formation to become tidally disrupted.  The arguments are based on analytic work as well as simulation data.  In this section, we discuss three consequences that clump disruption, as a general mechanism, can have on disk evolution.  Each subsection represents a proper study on its own, so our aim here is to present a foundation for subsequent work. The first topic, dust processing, is the most strongly connected to work presented in this manuscript, while the last topic is much more speculative.

\subsection{Dust Processing and Growth}

Figure 8 shows the dust mass in PC1 that is above temperatures of 1000 K, assuming no mixing or settling and a dust-to-gas ratio of $1/100$.  Mixing and an enhancing the dust-to-gas ratio by enrichment will allow for more solids to be processed.  \citet{helled_bodenheimer_2009} show that enrichment by planetesimals is unlikely for clumps that form on wide orbits, but this does not rule out enrichment at birth.  As discussed in section 3, fragmentation is most likely to form at the corotation of spiral waves.  These spirals will trap solids due to the pressure difference between the wave and the surrounding disk (Haghighipour \& Boss, 2002; Rice et al.~2004), ensuring that the dust is enriched where the disk is most likely to fragment. According to Figure 8, for roughly $5\times 10^4$ yr,  $>1 M_{\oplus}$ of dust will experience temperatures over 1000 K for our assumptions.   The figure also shows that a clump would need to survive for $\sim10^4$ yr before  significant high-temperature, high-pressure dust processing is expected to occur. Because C2 only survives for about 1500 yr, this particular clump is not expected to process dust thermally, unless it occurs in the disruption process itself.  However, there could have been substantial grain growth during this time, as meter-sized objects can grow in $\sim 1000$ yr for the conditions in the clump (e,g, Weidenschilling, 2000). Once the clump is disrupted, this material will be distributed into the disk. Because a clump on an eccentric orbit can experience multiple phases of mass loss due to changes in its Hill sphere, complete disruption is not necessarily required to liberate processed dust.  

For clumps that do experience high temperatures, their disruption, whether partial or entire, may be a mechanism for processing Calcium-Aluminium-Inclusions (CAIs) or CAI-like particles.  CAIs are composed of refractory minerals that are stable at temperatures above 1400 K (e.g., Scott 2007).  The advantage of this formation mechanism is that it can occur as soon as the disk begins to form; CAIs are among the oldest objects in the Solar System.   It is also a fallacy to assert that all of this material will be lost due to accretion onto the star.  GIs are known to transport material over large radii inward and outward, with significant mixing and stirring \citep[e.g.,][]{boss2004,boley_durisen2006}.  The retention of such material should be addressed, which we leave to future work, but we do not expect for all of this processed material to be lost.

\subsection{Core Formation}

 Clump disruption also may be capable of producing Earth-mass cores early in a disk's evolution.   These cores could provide a substantial head start for the core-accretion mechanism, and are not subject to the meter-barrier problem. To illustrate this possibility, we refer to \citet{helled_schubert2008}, who found that cores $\sim 1 M_{\oplus}$ can form in contracting proto-gas giants.  In their models, core formation was halted when the core temperature reached 1300 K, which begins silicate sublimation for the small grain sizes that they considered.  For clump masses $> 5 M_J$, they found that core formation was inhibited due to the fast evolution and high temperatures in the core.  We note that larger grains and highly refractory materials can survive beyond 1300 K \citep[see, e.g., Fig.~2 in][]{scott2007}, increasing the time (mass range) available for core formation.   

Any core that forms from a failed gas giant will not necessarily be lost due to type 1 migration, although it is not expected to stay at the disruption radius. \citet{laughlin_etal_2004} showed that the magnetorotational instability can lead to density fluctuations that will put cores on a random walk through the disk.  GIs should cause a similar behavior, but this needs to be explored in future work.  The cores that form in failed gas giants could become the first planetoids formed in planetary systems, and become seeds for the growth of gas and ice giants, or left as dwarf planets in an outer disk.   Earth-size objects in the Kuiper belt  would be consistent with clump disruption.

\subsection{FU Ori}
A massive clump will remain susceptible to tidal disruption for $10^4$ to $10^5$ yr.  During this time, transport of a clump into the innermost regions of the disk ($r<1$ AU) may become possible through clump scattering and/or disk torques.  This topic warrants further study, but we only emphasize this as a real possibility during the early phases of disk evolution while heavy infall is present on the disk \citep{vorobyov_basu2006,boley2009}.  For a $\sim 0.3 M_{\odot}$ star, the Hill radius of a 3 $M_J$ object is $\sim 0.3$ AU at a disk radius of 2 AU.  For our simple contraction model, we find that the radius of PC1 is comparable to this size just as it hits $T_c\sim 2000$ K.  Clump disruption inside a few AU can suddenly supply the inner disk with mass, which may lead to a thermal instability \citep{bell_lin1994}, and consequently, an FU Orionis outburst.  Note that we do not require the clump to make it all the way to the star.   For a very eccentric scattering event, such that the clump arrives at perihelion at nearly the escape speed, a Jupiter-mass fragment around a 0.5 M$_{\odot}$ primary will cause a maximum change in the primary's radial velocity of about 60 m/s.  Unfortunately, the radial velocity drifts of, e.g., FU Ori have only been constrained to 300 m/s (Petrov \& Herbig 2008).  Larger clump/primary ratios and smaller perihelion passages may allow for an observable radial drift in other systems. 

\section{Conclusions}

We have demonstrated that initial clump masses are expected to be in the gas giant regime. Our analytic mass estimates are consistent with hydrodynamics simulations with radiative cooling.  These clumps will have significant angular momentum, suggesting that disks $\sim$ AU in radius should form after dissociative collapse; the initial clump mass represents what will eventually become a core-disk system.  Even when mass is assumed to be accreting at its maximum rate, at least with our analytic estimates, 1 M$_J$ clumps are only expected to grow to masses $\sim 10$ M$_J$ before H$_2$ dissociation causes rapid collapse, consistent with known masses for the planets in HR 8799 \citep{marois_etal_2008}.  Further growth of the system may be regulated by the circumplanetary/brown dwarf disk due to the angular momentum of the new material entering the Hill sphere. Clump $\tau_1$ contraction timescales, even for several $M_J$, will be $10^4$ to $10^5$ yr, giving sufficient time for clumps to be transported into the inner disk and to be tidally destroyed.  Clumps will have very different environments from the typical conditions in the outer disk, and they represent factories for processing dust and building large solid bodies.  Clump disruption therefore represents a mechanism for processing dust, modifying grain growth, and building large, possibly Earth-mass, objects during the first stages of disk formation and evolution. 

%% Using an acknowledgements command is not in the Elsevier template,
%% but it can be used.
\ack
We would like to thank both referees for very useful critiques that improved this manuscript.  A.C.B's contribution was supported by the University of Zurich, an SNF grant, and the University of Florida.  T.H.'s was supported by ETH, and L.M.'s contribution by ETH, the University of Zurich, and an SNF grant. R.H.D.'s contribution was supported by NASA Origins Grant NNX08AK36G.  The simulation presented here was run on the zbox machines, maintained by ITP, University of Zurich.

\label{lastpage}

% Bibliographic references with the natbib package:
% Parenthetical: \citep{Bai92} produces (Bailyn 1992).
% Textual: \citet{Bai95} produces Bailyn et al. (1995).
% An affix and part of a reference:
%   \citep[e.g.][Ch. 2]{Bar76}
%   produces (e.g. Barnes et al. 1976, Ch. 2).-

%\newcommand{\chondrites}{2005, in ASP Conf.~Ser.~341, Chondrites and the protoplanetary disk}

\bibliography{bibliography.bib}

%% Use the plainnat style for ``Icarus'' mode to display DOI numbers
%% among other things.  However, revert to the Elsevier elsart-harv
%% mode for ``Elsevier'' mode.
\bibliographystyle{plainnat}
% \bibliographystyle{elsart-harv}

%% --Tables-- 

\clearpage	% Make sure things don't run together.
\begin{table}
\begin{center}
\begin{tabular}{ l l l l l l }
Clump & Mass ($M_J$) & $R_{\rm sphere}$ (AU) & $a$ (AU) & J ($\rm cm^2 s^{-1}$) & Formation (yr) \\\hline
C1 & 3.3 & 2.9 & 80 & 2.3(18) & 300\\
C2 &  1.7 & 2.3 & 65 & 1.1(18) & 750 \\
C3 & 0.87 & 1.8 &  85 & 3.4(17) &  970\\
\end{tabular}
\caption{Initial clump parameters from the simulation SPHSIM.  The value $R_{\rm sphere}$ is the radius that a sphere would have for the same volume as the clump.  The formation radius is indicated by $a$. The masses and $J$s for C1 and C2 are consistent with our estimates in sections 3 and 4.  C3 seems to form as a result of an interaction between an existing clump's spiral wake and a wave, and is consistent with delayed fragmentation.  The last column indicates the formation time after the start of SPHSIM. \label{lasttable}}
\end{center}
\end{table}

\clearpage

%% --Figures-- %%

\begin{figure}[p!]
\begin{center}
\includegraphics[width=8cm,angle=-90]{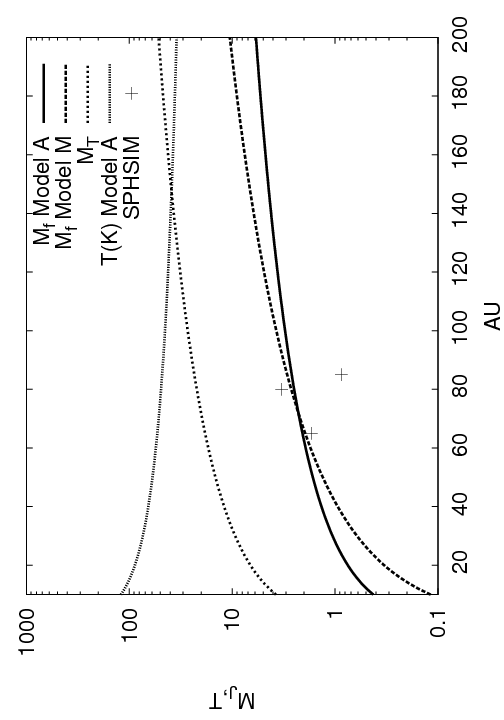}
\caption{Mass estimates calculated by assuming the Toomre mass $M_T=\pi \Sigma (\lambda_T/2)^2$ and by calculating $M_f$ as described in the text. The curve associated with Model M represents the fragmentation mass that one would expect in the simulation.  The curve labeled ``T(K) Model A'' gives the temperature profile for Model A.  The simulation data (crosses) are in good agreement with the $M_f$ estimates.}
\end{center}
\end{figure}

\clearpage

\begin{figure}[p!]
\begin{center}
\includegraphics[width=6cm]{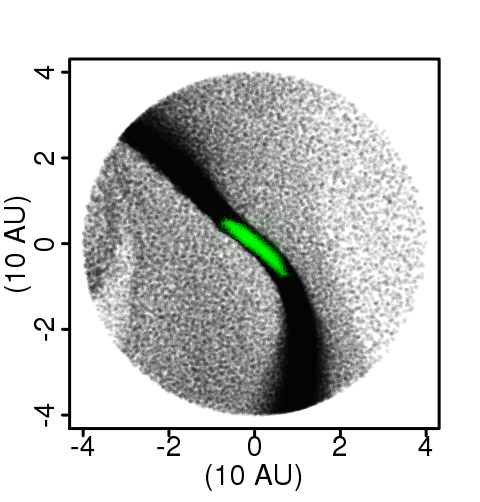}\includegraphics[width=6cm]{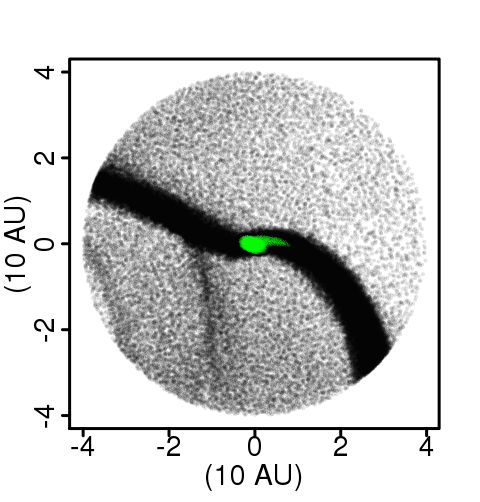}
\caption{The particle distribution for C1 just before and after fragmentation.  Green represents particles that are included in the initial mass given in Table 1.}
\end{center}
\end{figure}

\clearpage

\begin{figure}[p!]
\begin{center}
\includegraphics[width=8cm,angle=-90]{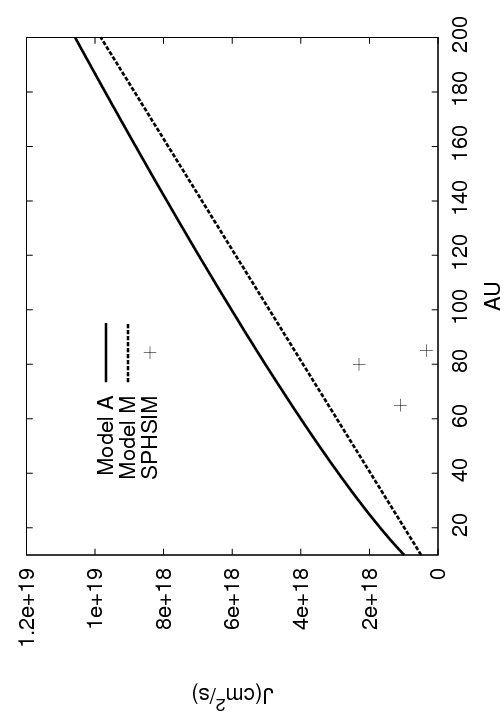}
\caption{Specific angular momenta for Models A and M.  Model M shows what we expect to see in SPHSIM.  The simulation data are indicated by crosses.}
\end{center}
\end{figure}

\clearpage

\begin{figure}[p!]
\begin{center}
\includegraphics[width=8cm,angle=-90]{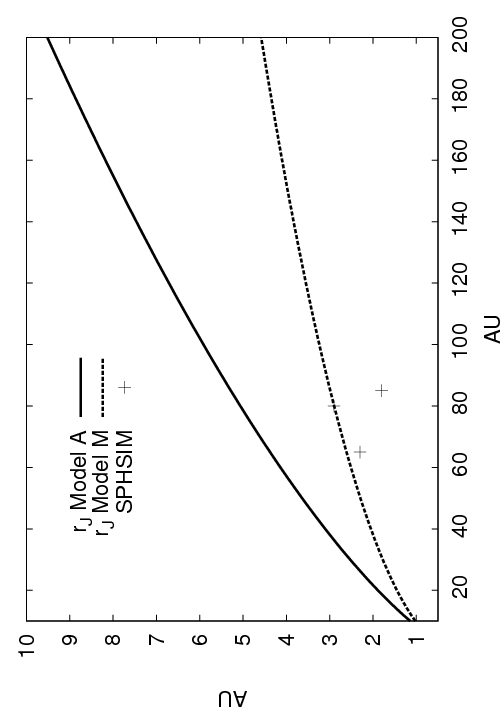}
\caption{Angular momentum radius $r_J$  for or Models A and M.  Model M shows what we expect to see in SPHSIM. The simulation data for $R_{\rm sphere}$ are indicated by crosses.  Because the $J$s for the clumps are lower than expected by a factor of order unity, the correspondence between $R_{\rm sphere}$ and $r_J$ may be due to thermal support. }
\end{center}
\end{figure}

\clearpage

\begin{figure}[p!]
\begin{center}
\includegraphics[width=8cm]{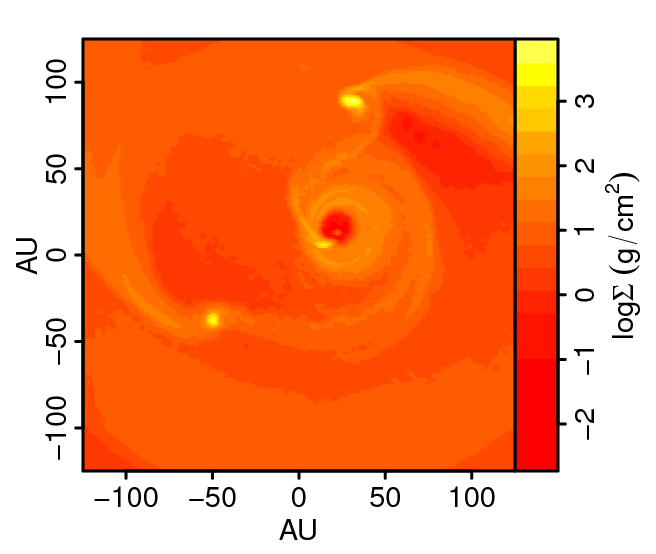}
\caption{Surface density snapshot from the simulation SPHSIM.  The innermost clump (now quite elongated at x,y$\approx$20,5 AU) is being tidally destroyed, and does not survive for more than 1/4 more of an orbit.  C1, near the top of the snapshot, is becoming bar unstable.}  
\end{center}
\end{figure}

\clearpage

\begin{figure}[p!]
\begin{center}
\includegraphics[width=8cm,angle=-90]{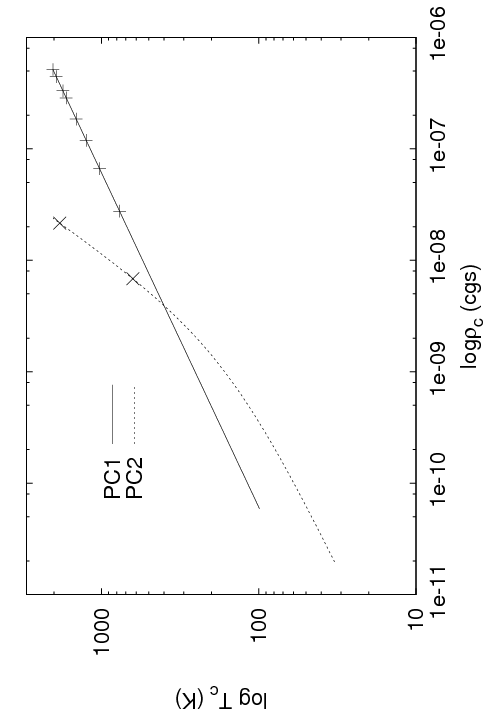}
\caption{The $\rho -T$ plane for the contracting polytrope models.  Symbols indicate roughly $10^4$ yr intervals.  Once the central temperature reaches 2000 K, rapid collapse should occur. PC1 is a 3 M$_J$ clump contracting in isolation.  PC2 starts at 1 M$_J$ and accretes mass as fast as gas can flow into the clump's Hill sphere.}
\end{center}
\end{figure}

\clearpage

\begin{figure}[p!]
\begin{center}
\includegraphics[width=8cm,angle=-90]{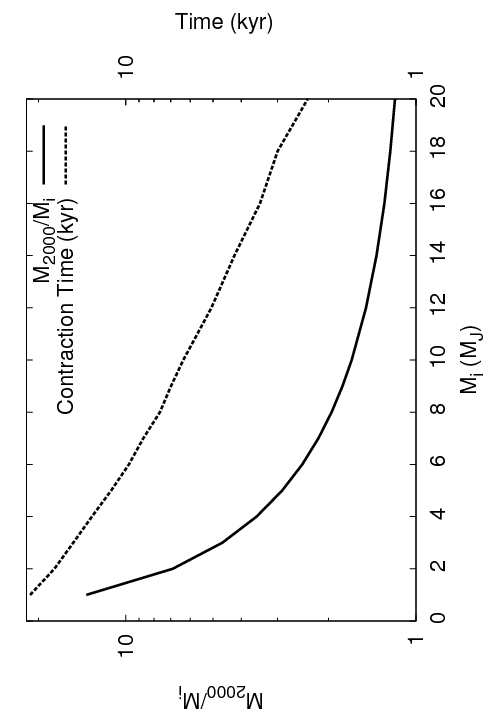}
\caption{The ratio of $M_{2000}$ to the initial fragment mass $M_i$, where $M_{2000}$ is the mass of the clump when the central temperature reaches 2000 K, leading to rapid collapse.  The contraction time in kyr is shown with the dashed line, and is on the same scale as the mass ratio curve.  Even with accretion, clumps with initial masses below $\sim 6$ M$_{J}$ remain susceptible to disruption for $10^4$ yr.  For masses greater than roughly 8 M$_J$, the clump collapses before the mass can be doubled.  By $M_i=20$ M$_J$, accretion is marginalized.}
\end{center}
\end{figure}

\clearpage

\begin{figure}[h*]
\begin{center}
\includegraphics[width=8cm,angle=-90]{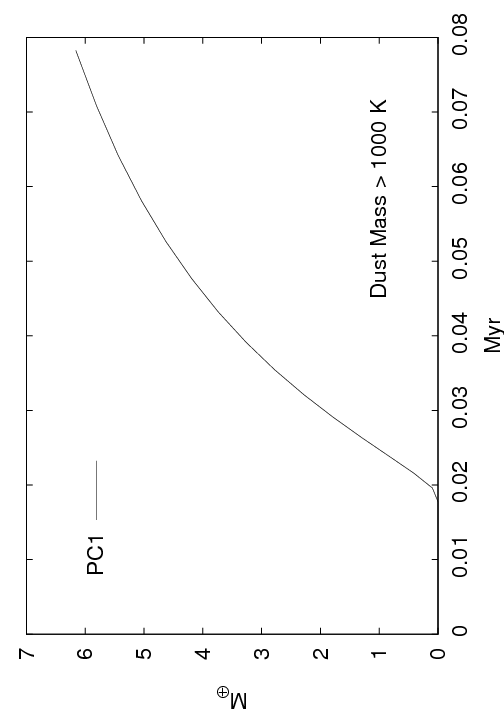}
\caption{Dust mass in regions with $T>1000$ K.  We have assumed no settling or mixing and a dust-to-gas ratio of 1/100.}
\end{center}
\end{figure}

\label{lastfig}

\end{document}